\documentclass[aip,reprint]{revtex4-1}
\pdfoutput=1
\usepackage{graphicx}
\usepackage{amsmath}
\usepackage{amssymb}
\usepackage{color}
\definecolor{amber}{rgb}{1,0.49,0} 
\usepackage{multirow}
\usepackage{siunitx}
\newcommand\fref[1]{Figure \ref{#1}}

\begin{document}


\title{
Tipping detection using climate networks}



\author{Laure Moinat}
\email[]{laure.moinat@unige.ch}
\author{J\'er\^ome Kasparian}
\author{Maura Brunetti}
\affiliation{Group of Applied Physics and Institute for Environmental Sciences, University of Geneva, 66 Bd Carl-Vogt, CH-1211 Geneva 4, Switzerland}


\date{\today}





\begin{abstract}
The development of robust Early Warning Signals (EWS)  is necessary to quantify the risk of crossing tipping points in the present-day climate change. Classically, EWS are statistical measures based on time series of climate state variables, without exploiting their spatial distribution. However, spatial information is crucial to identify the starting location of a transition process, and can be directly inferred by satellite observations. By using complex networks constructed from several climate variables on the numerical grid of climate simulations, we seek for network properties that can serve as EWS when approaching a state transition. We show that network indicators such as the normalized degree, the average length distance and the betweenness centrality are capable of detecting tipping points at the global scale, as obtained by the MIT general circulation model in a coupled-aquaplanet configuration for CO$_2$ concentration-driven simulations. The applicability of such indicators as EWS is assessed and compared to traditional methods. We also analyse the ability of climate networks to  identify  nonlinear dynamical patterns.
\end{abstract}


\maketitle 


{\bf Present-day anthropogenic CO$_2$ forcing causes the Earth climate to vary in an extreme manner, which can lead to abrupt changes at a local or even global scale. Such abrupt changes in the climate are known as tipping processes. Detecting and anticipating them is crucial in our efforts to limit the consequences of climate change. Several techniques have been developed using time series analysis and have given rise to early warning signals (EWS) of tipping processes. However, these methods do not consider the spatial evolution of climate change. To combine both temporal and spatial information, we consider the time evolution of climate networks. The idea is to grid the Earth's surface with networks, thus capturing the spatial propagation of change. We test this method 
on a global-scale tipping using climate simulations. By including the information of the spatial distribution, these networks are able to serve as better EWS compared to classical methods and to reveal the regions where the tipping develops.}
\section{Introduction}
\label{sec:intro}

Crossing tipping thresholds in the present-day climate change is raising increasing concern~\citep{steffen2018,IPCC} as the global mean temperature is getting closer to exceeding the Paris agreement range. Tipping elements have been identified with impacts at the regional and global scale~\citep{lenton2008,lenton2019,McKay2022,wang2023}. The interactions among these elements can likely induce tipping cascades with the risk of accelerating irreversible changes in the climate system~\citep{dekker2018,brovkin2021,niko2024}. In this context, it becomes urgent to develop robust early warning signals (EWS) for detecting the approach of tipping points. 

In the past, a variety of methods has been explored to obtain EWS. The classical approach is based on the bifurcation theory and the phenomenon of `critical slowing down'~\citep{Dakos2008,scheffer2009}, where changes in the variance, the lag-1 autocorrelation or high-order momenta of the system's time series are used as EWS~\citep{Carpenter2006,Guttal2008}. However, several other techniques have been proposed in the literature such as estimating fluctuations decay~\citep{livina2007}, frequency change~\citep{Williamson2015} or flickering~\citep{dakos2013flickering} in noisy time series, and basin entropy~\citep{entropy} near Hopf bifurcations. Moreover, physics-based indicators have been also tested for particular tipping elements, such as the freshwater transport for the tipping of the Atlantic Meridional Overturning Circulation (AMOC)~\citep{deVries2005,vanwesten2024}. 

In general, the above mentioned indicators provide information on the temporal development of a tipping process, without exploiting its spatial structure. However, critical slowing down arises when a system becomes increasingly sensitive to external perturbations and not necessarily near a catastrophic shift~\cite{fauxpositifs}. This shows the need of including the information from the spatial processes during tipping events to have robust EWS and reduce the number of false positives~\citep{dakos2010}. Spatial methods include the edge detectors in spatiotemporal 
data~\citep{bathiany2020edge} and indicators based on climate networks~\citep{bookNetworks2019}. These methods become particularly relevant in the current era of huge amounts of data coming from satellite observations with high spatiotemporal coverage~\cite{lenton2024}. Network indicators have been successfully applied for studying climate subsystems, like the AMOC tipping~\citep{vanderMheen2013} and El Ni\~no southern oscillation~\citep{Donges_2009b,Radebach_2013}.  

Here, we investigate how network indicators behave as EWS during a tipping point at the planetary scale. 
Tipping between climatic attractors is simulated with the MIT general circulation model in a coupled-aquaplanet configuration~\citep{Brunetti_2019} by increasing or decreasing the atmospheric CO$_2$ content at different rates. 
Using the Python package \texttt{pyUnicorn}~\cite{Donges_2015}, we construct  networks based on linear and nonlinear spatial correlations of time series at each point of the numerical grid. We test several indicators that characterise the network topology on the local, mesoscopic and global scales~\citep{Donges_2009}. We apply them to various networks, constructed from different climatic state variables, spanning the whole planet or only latitudinal strips with several time window sizes. 
The aim of this detailed analysis is to understand the ability of network indicators to predict the occurrence of global tipping points and to unveil nonlinear backbone structures in the climate system.

\section{Method} 
\label{sec:methods}

In this section, we will briefly describe the main characteristics of the climate networks and the simulations data set we are going to use. To generate the climate networks and calculate the network indicators we use the {\tt pyUnicorn} package in python~\citep{Donges_2015}, as detailed in App.~A.  

\subsection{Climate networks} 

A network is a set of points that  interact with each other during a certain period of time. In this work, we will consider only 
undirected networks, i.~e., without considering the direction of interrelationships. 

The vertices (nodes) $p = 1,2, ..., N$ of a climate network are the spatial grid points of an underlying global climate data set, which can be derived from either observations or simulations.
Specifically, we take as vertices the points on a longitude-latitude grid.
Area-weighted indicators are then used to avoid the heterogeneous spatial distribution of vertices to bias the network topological proprieties~\citep{Donges_2009,Heitzig2012,Donges_2015}. 

Edges $E$ (links) between pairs of vertices $(p,q)$ are defined depending on the degree of statistical interdependence between the corresponding pairs of time series taken from the climate data sets (pressure, temperatures, precipitations, etc.). The number of edges varies from vertex to vertex and can evolve in time~\citep{Donges_2009,Radebach_2013}. 
The statistical interdependence depends on a threshold value, which is a crucial choice as it defines the existence of an edge between two vertices. Indeed, for the same amount of vertices, using a different threshold gives rise to a different edge density, given by: 
\begin{equation}
\rho = \frac{E}{\binom{N}{2}} 
\end{equation}
where $E$ is the number of edges and the binomial coefficient $\binom{N}{2} = N(N-1)/2$ is the number of potential edges in the network.
The resulting connections between vertices can be represented by 
a binary symmetric matrix $A$ called the {\it adjacency matrix}: 
\begin{equation}
  A_{pq} =
    \begin{cases}
      0 & \text{if $(p,q) \notin E$}\\
      1 & \text{if $(p,q)\in E$}\\
    \end{cases}       
\end{equation}
Each non-zero element of $A_{pq}$ represents an edge of the network for the $(p,q)$ pair~\citep{Zou_2019}. For a network containing $N$ vertices, the corresponding adjacency matrix will be a $N\times N$ matrix \citep{Barreiro_Networks_in_climate}. 

In this work, we will consider two statistical ways of generating a complex network: the Pearson correlation coefficient (PCC) and the mutual information (MI)~\citep{Donges_2009}. The reason for this choice is that PCC is commonly used in the literature, but reveals only linear correlations between time series. In contrast, MI has the ability to reveal non-linear correlations, as shown in~\citet{Donges_2009}.

PCC measures the similarity of two times series $X, Y$, sampled as a function of the time shift $\tau$, and is defined as: 
\begin{equation}
    \rho_{X,Y}(\tau) = \frac{\text{cov}(X(t), Y(t+\tau))}{\sqrt{\sigma_X^2\sigma_Y^2}} 
\end{equation}
where $\sigma_X^2$ and $\sigma_Y^2$ are the variances and $\text{cov}(X(t), Y(t+\tau))$ is the covariance of the two time series. $\rho_{X,Y}(\tau)$ has the propriety that $\rho_{X,Y}(\tau) = \rho_{Y,X}(-\tau)$, thus it is symmetric \citep{Barreiro_Networks_in_climate} and bounded between $-1$ and $1$. 

MI measures the information gained about a time series of one variable given the knowledge on another time series. If we consider two time series $X$ and $Y$, it is defined as:
\begin{equation}
    MI_{X,Y} = \sum_{x}\sum_{y} p(x,y) \log \left( \frac{p(x,y)}{p(x)p(y)} \right)
\end{equation}
where $p(x)$ and $p(y)$ are the probability distributions and $p(x,y)$ the joint one \citep{Barreiro_Networks_in_climate}. 
MI is always  positive. 

If two time series are heavily correlated linearly, the $M I$ value will be large and $|\rho_{X,Y}|$  will be close to $1$. In the case of a strong non-linear correlation between $X$ and $Y$, $MI_{X,Y}$ will be large, while $\rho_{X,Y}$ small. Therefore, by comparing them it is possible to identify the contribution of non-linear dynamics in a complex network. 

\subsection{Networks indicators}
\label{subsec:indicators}

To study the topological proprieties of the network, several indicators are available in the literature. The first indicator used in this work is the {\it normalized degree centrality} $k_v$, which is defined in terms of the adjacency matrix~\citep{Freeman_1978}:
\begin{equation}
    k_v = \frac{1}{N}\sum_{i=1}^{N} A_{vi}
\end{equation}
This indicator gives the number of correlated relations per vertices. 
High-degree vertices are identified as super-nodes. 

The next indicator is the {\it local clustering coefficient} of a vertex $v$:
\begin{equation}
    C_v = \frac{E(\Gamma_v)}{\binom{k_v}{2}}
\end{equation}
where $\Gamma_v$ is the number of first neighbours of $v$, $E(\Gamma_v)$ the number of edges connecting the first neighbours, and the binomial coefficient $\binom{k_v}{2} = k_v(k_v - 1)/2$ gives the maximum number of edges in $\Gamma_v$. 

From this indicator  the {\it global clustering coefficient} can be obtained taking the mean over all vertices:
\begin{equation}
    C = \langle C_v \rangle_v
\end{equation}
This indicator gives the probability that, for a random chosen vertex, its first neighbours are also connected between each other \citep{Strogatz_1999}. 

Another indicator is the {\it average path length} (or {\it average length distance}), that is, the topological distance between pairs of vertices:
\begin{equation}
    \mathcal{L} = \frac{1}{\binom{N}{2}}\sum_{i<j}d_{ij}
\end{equation}
where $d_{ij}$ is the minimum number of edges that need to be crossed to travel from vertex $i$ to vertex $j$. Disconnected pairs are not included. 

Finally, the {\it betweenness centrality} of a vertex is defined as:
\begin{equation}
    B_i = \frac{\sum_{jk\not = i}\sigma_{jk}(i)}{\sigma_{jk}}
\end{equation}
where $\sigma_{jk}$ is the number of node pairs and $\sigma_{jk}(i)$ is the number of shortest paths between $j$ and $k$ that include vertex $i$.
It has been shown that $B_i$ is significantly affected by the presence of nonlinear processes  when mutual information is used to generate the network, and therefore unveils non-linear edges~\citep{Donges_2009}. 

All  these indicators use different ranges of information. The {\it degree centrality} uses the local information from the vertices that are connected between each other. Whereas the {\it local/global clustering coefficients} provide information on a mesoscopic scale, since they depend on the information from their connected neighbours. Finally, the {\it average path length} and the {\it betweenness centrality} use global topological information as they consider the shortest path among different connected pairs, which can span the whole network~\citep{Donges_2009}.

\subsection{Climate simulations}
\label{sec:sims}

\begin{figure}   
\includegraphics[width=0.49\textwidth, keepaspectratio]{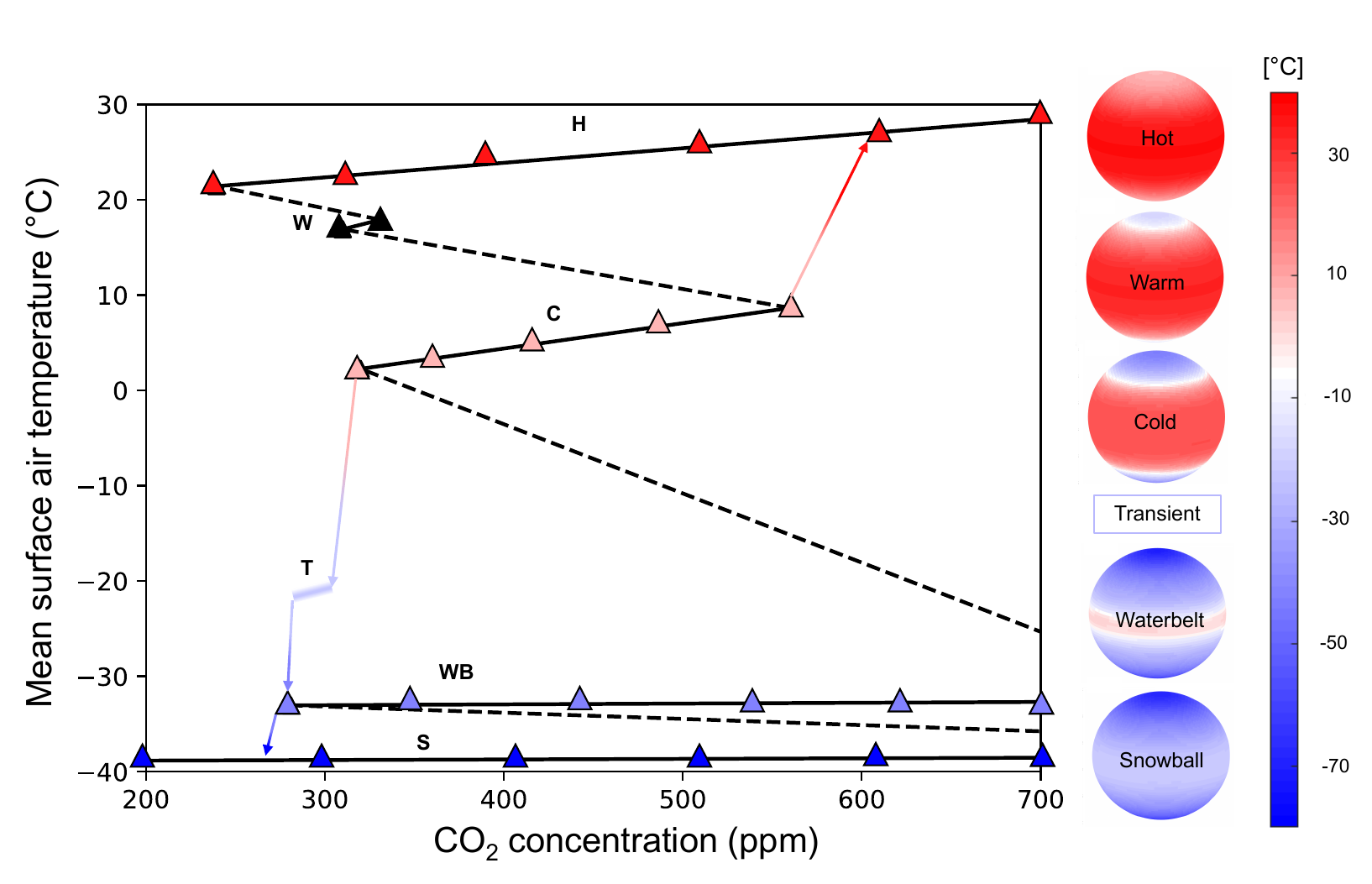}
\caption{Bifurcation diagram for the coupled aquaplanet with the five steady states, the transient state and the two transitions C-H and C-S highlighted. Dashed lines correspond to qualitative sketches of unstable branches.}
\label{Diagram_bifurc}
\end{figure} 

Using the MIT general circulation model (MITgcm)~\citep{marshall_finite-volume_1997,marshall_hydrostatic_1997,adcroft_implementation_2004,marshall2004}, we consider the same coupled-aquaplanet configuration analysed in~\citet{Brunetti_2019,Ragon_2022,Ragon_2023}, i.~e., a planet entirely covered by a 3000~m-deep ocean where full dynamical couplings between ocean and atmosphere are included. In this configuration, five climatic attractors (which are stable over a multi-millennial time scale) have been found~\citep{Brunetti_2019}, denoted as snowball (S), waterbelt (WB), cold (C), warm (W) and hot (H) states as shown in the bifurcation diagram in Fig.~\ref{Diagram_bifurc}. 
Here, we consider the atmospheric CO$_2$ content as a forcing parameter. We have completed the bifurcation diagram in terms of the equilibrium surface air temperature (SAT), partially shown in the Supplemental Material of~\citet{Ragon_2023}.
This kind of diagrams allows one to know where the bifurcation-induced tipping points occur, the range of the stable branches and the regions of multistability, revealing the core dynamical structure of the climate system.  

We investigate here the transitions at the global scale from one attractor to the other. 
We are going to focus on two types of transition: one from the cold state to the hot state (C-H), and one from the cold state to the snowball state (C-S), with intermediate transitions to a transient state around $-20~^\circ$C (which is stable only over a timescale of hundred years, as verified in~\citet{Ragon_2023}) and to the waterbelt state, as shown in Fig.~\ref{Diagram_bifurc}. 
The reverse transition from the hot state to the cold state is not possible in our configuration because of the hysteresis path between these steady states. 
Indeed, from the bifurcation diagram in Fig.~\ref{Diagram_bifurc}, the climate system on the hot-state branch can only tip towards the snowball state by gradually reducing the forcing. Therefore, we choose to look at a transition starting from the same cold attractor as the upward transition, but going to colder states.
With climate change we are expecting an upward transition (in terms of global temperature), nevertheless it is important to verify which is the response of climate networks to both upward and downward transitions. 
Moreover, the presence of an intermediate transient state between the cold and waterbelt states enables us to estimate the network sensitivity. 

The tipping is induced by varying the atmospheric CO$_2$ content. We have performed several simulations using different forcing rates, ranging from 0.1~ppm/yr (which is sufficiently low to guarantee that the system relaxes toward the attractor at each step) to 2~ppm/yr (which is the present-day rate).

We consider three state variables with yearly frequency that provide different information on the climate dynamics. SAT is treated in detail in the main part of this article, while specific humidity (SH) and cloud cover (CC) can be found in the appendix for the results part. SAT and SH are prognostic variables in MITgcm, directly linked to the energy and water cycles, respectively, whereas the CC fraction is a diagnostic variable obtained from temperature, pressure and specific humidity~\cite{molteni_atmospheric_2003}.

\subsection{Determination of the tipping time interval}

To fully describe the transition we  define three main intervals for the temporal evolution of a climatic state variable: on the initial attractor, during the tipping process and on the final attractor. 
They are marked by the start $t_1$ and the end $t_2$ of the tipping, and can be precisely defined from the simulations by the following two criteria~\cite{Ragon_2023}: (i) the ocean surface radiation imbalance is lower than 0.5~Wm$^{-2}$ in absolute value;  (ii) under forcing conditions, the system does not shift monotonically in time toward other values.  
When one of these criteria is not satisfied, the system enters in the tipping (or transition) interval. 
The precise definition of the tipping interval is important regarding our analysis on detecting tipping points and early warning signals' robustness. The above analysis is repeated for each state variable in each simulation, and the values of $t_1$ and $t_2$ are reported in Table~\ref{tipping_C2H} for the C-H transition and in Table~\ref{tipping_C2S} for the C-S transition. 

\begin{table}[]
    \centering
    \begin{tabular}{c|cc|cc}
     C-H &  \multicolumn{2}{c|}{ 0.1~ppm/yr} & \multicolumn{2}{c}{ 2~ppm/yr}  \\
     variable & $t_1$ [yr] & $t_2$ [yr] & $t_1$ [yr] & $t_2$ [yr]  \\
     \hline
SAT & 2400 & 3400 & 200 & 700  \\
CC & 2300 & 3300 & 150 & 700  \\
Humidity & 2400 & 3300 & 200 & 600  
    \end{tabular}
    \caption{Values of the start $t_1$ and the end $t_2$ of the tipping interval for different forcing rates in the C-H transition.}
    \label{tipping_C2H}
\end{table}

\begin{table}[]
    \centering
    \begin{tabular}{c|cc|cc}
     C-S &  \multicolumn{2}{c|}{ 0.1~ppm/yr} & \multicolumn{2}{c}{ 0.4~ppm/yr}  \\
     variable & $t_1$ [yr] & $t_2$ [yr] & $t_1$ [yr] & $t_2$ [yr]  \\
     \hline
SAT & 300 & 1750 & 150 & NA  \\
CC & 300 & 1750 & 150 & NA  \\
Humidity & 300 & 1650 & 150 & NA  
    \end{tabular}
    \caption{Values of the start $t_1$ and the end $t_2$ of the tipping interval for different forcing rates in the C-S transition. }
    \label{tipping_C2S}
\end{table}

Contrary to C-H, C-S shows the presence of an intermediate transient attractor, which has been analysed in detail in~\citet{Ragon_2023}. In addition, the ocean surface imbalance displays large fluctuations until around 1000~yr. This strong difference in the evolution profile compared to the C-H transition allows us to test the robustness of the networks and their capability of detecting fluctuations in the tipping process.

\subsection{Threshold value for edge definition}
\label{subsec:threshold}

The choice of the threshold above which pairs of vertices are considered to be connected is a critical step in the network construction. For PCC, we tested threshold values from 0.1 to 0.98 with increments of 0.02. For MI, we tested values from 0.005 to 1.5 with increments of 0.005. In both cases, we performed the test once while the system is on the attractor and once during the tipping process.

\begin{figure}   
\includegraphics[width=0.48\textwidth, keepaspectratio]{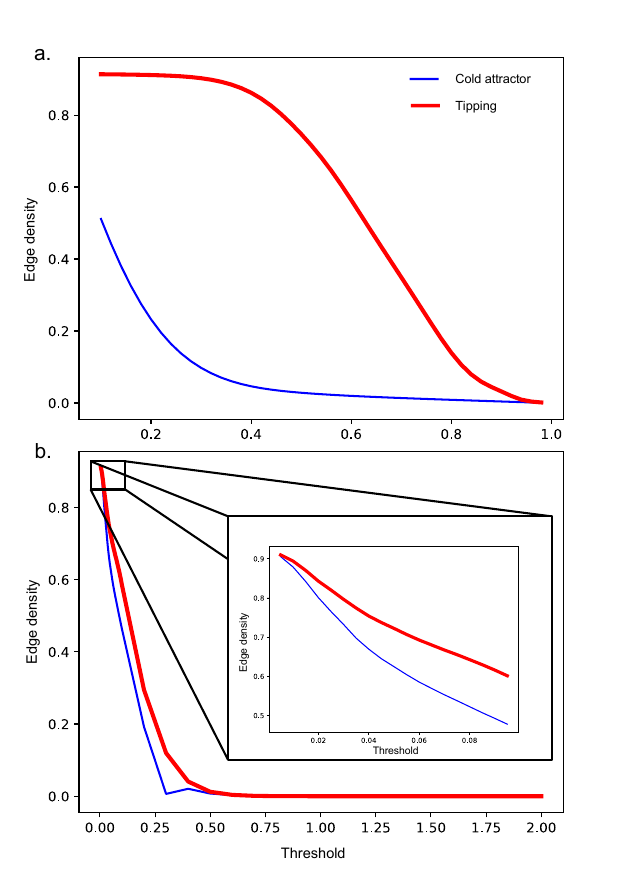}
\caption{Evolution of the edge density against the threshold value using a. Pearson correlation coefficient (PCC) and b. mutual information (MI) on the attractor  (on the cold state at $t=2000$~yr during the C-H transition shown in Fig.~\ref{Figure_explique}) and during the tipping (at $t=2850$~yr). }
\label{Threshold_final}
\end{figure} 

As shown in \fref{Threshold_final}.a, during the tipping there is a saturating effect for PCC if we choose a value lower than 0.4 (red thick curve). Therefore, if we take a threshold below this value, the evolution of the edge density of the PCC is saturated, and the network dynamics cannot be properly studied during the tipping.
Therefore, the threshold in PCC needs to be carefully chosen. 
To ensure that our analysis is not affected by the saturating effect while still considering as many edges as possible, we use a threshold value of 0.5 for PCC. 
In contrast,  there is no such saturating effect in MI, as shown in \fref{Threshold_final}.b. 

If we want to compare the two methods of network construction, it is more relevant to use the same edge density and not the same thresholds as they are not commensurate~\citep{Donges_2009b}. The edges that are not present in the PCC adjacency  matrix but are found in the MI matrix contribute to the non-linear dynamics of the complex network~\citep{Donges_2009}.  

We have also examined what happens when the edge density is fixed instead of the threshold. In this case, even though the overall behaviour during the tipping process is similar to what is observed with a fixed threshold, it is difficult to track the evolution of all the spatial regions because only the strongest and most dominant edges exist, and therefore other weaker correlations are neglected. 

\section{Results}

\begin{figure*}[t!]   
\includegraphics[width=1.0\textwidth, keepaspectratio]{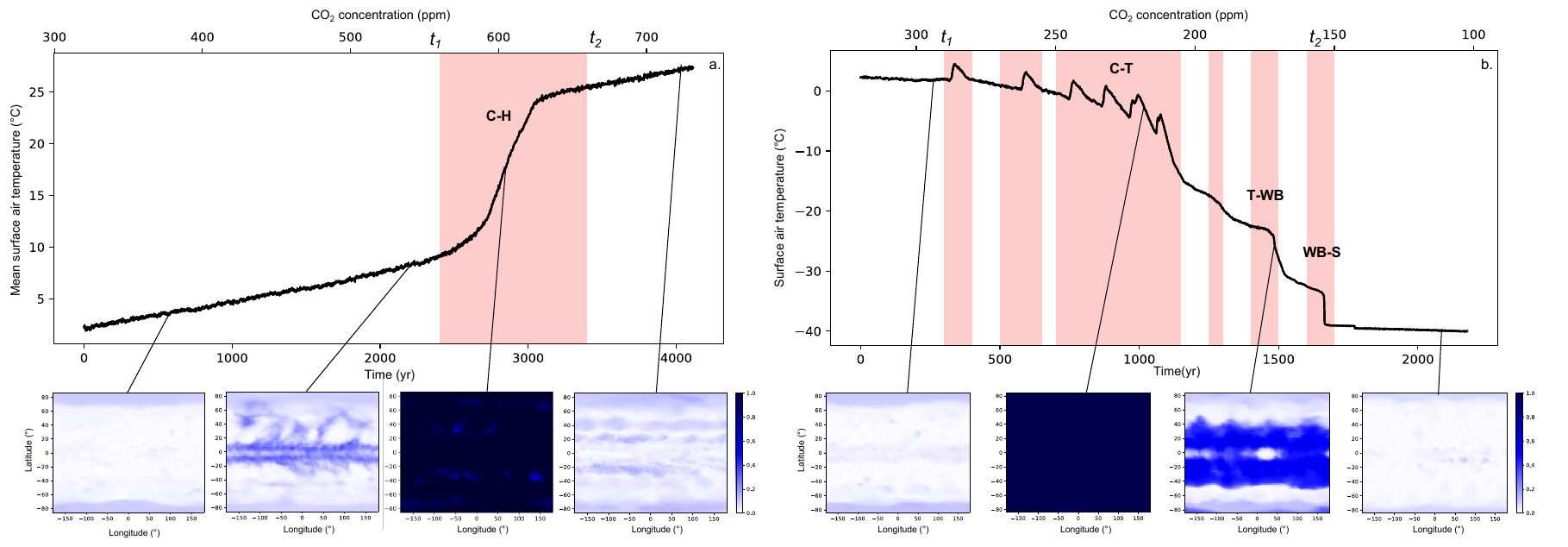}
\caption{Temporal evolution of the mean surface air temperature during the transition from the cold to the hot state (C-H, left) and the cold to the snowball state (C-S, right), passing through the cold to transient (C-T), transient to waterbelt (T-WB), and waterbelt to snowball (WB-S) tipping intervals, when the atmospheric CO$_2$ content increases or decreases at a rate of 0.1~ppm/yr. For selected time slices on the attractors and during the tipping, the map of the normalized degree centrality is shown in the bottom panels. }
\label{Figure_explique}
\end{figure*} 

As detailed in the Method section, we consider how climate networks behave during an upward transition going from the cold to the hot state (C-H) and a downward transition going from the cold state to the snowball (C-S). 

\subsection{General behaviour}

The evolution of SAT during the C-H transition is shown in the left panel of Fig.~\ref{Figure_explique}. The initial state is on the cold attractor at an atmospheric  CO$_2$ content of 320~ppm. By increasing it by 0.1~ppm per year, the system evolves along the stable branch of the cold attractor until the tipping starts. At this point, SAT increases much faster, finally arriving at the hot attractor.   
During this evolution, the climate network evolves accordingly. As an example, the bottom panels of Fig.~\ref{Figure_explique} show the behaviour of the normalized degree centrality in a PCC network. We can see that during the tipping process the number of spatial correlations  increases~\citep{dakos2010}, as observed in phase transitions of thermodynamic physical systems~\citep{bookNetworks2019}. This is predicted by the {\it slaving principle}~\citep{Haken2004}, where at a bifurcation all the dynamical modes relax and adapt to the slowly evolving one, thus giving rise to spatial correlations that extend to the whole system. 

Before and after the tipping process, the network describes the internal variability of the attractor dynamics. In the cold state, the main correlations occur in the polar regions, while in the hot state the patterns are more uniform and also involve the mid-latitude regions, in agreement with the principal component analisys reported in~\citet{Brunetti_2019}. 

In the following, we will characterize in detail the ability of climate networks on the basis on both their spatial and temporal information to describe this type of global-scale tipping.

\subsection{Sensitivity tests}

As a preliminary step, we need to set some free parameters and evaluate their effects on the network dynamics. 
In particular, we  
consider the impact of changing the window size, as well as the discretization and intensity of the forcing rate. 
Unless otherwise specified, the tests are made using the C-H transition with a forcing rate of 0.1~ppm/yr and SAT as climate state variable. 

\subsubsection{Time window size}

\begin{figure*}   
\includegraphics[width=1.0\textwidth, keepaspectratio]{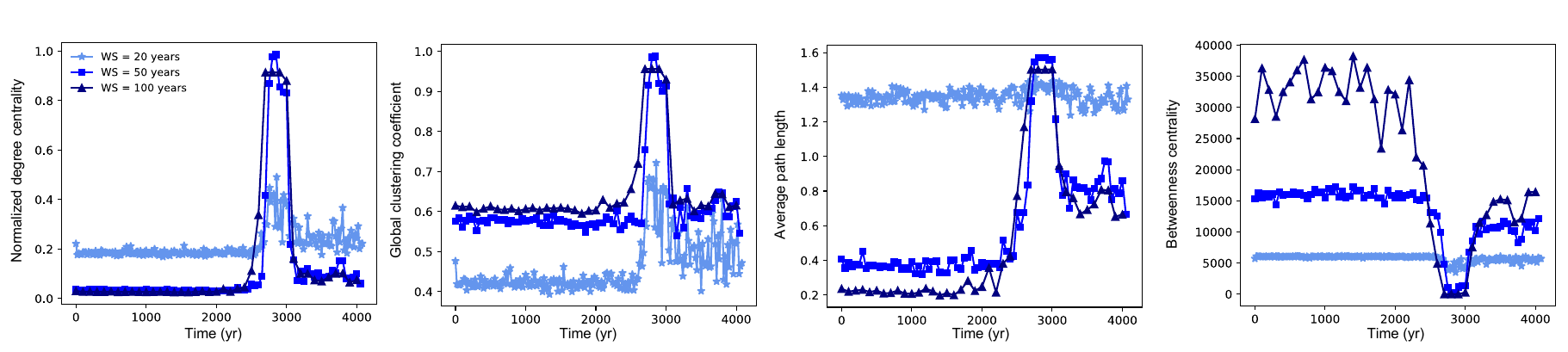}
\caption{Temporal evolution of four network indicators constructed with three different window sizes (WS).}
\label{Window_size}
\end{figure*} 

Testing this free parameter is crucial for the reliability of the network outcomes analysis. Varying the window size affects the amount of information from the time series, and therefore the significance of the network indicators~\cite{Wu2024}.   
The task is to find a compromise between maximizing the amount of information at each time step  on one side, and a high temporal resolution on the other side, with a sufficient number of points  in the temporal evolution of the network in order to provide a proper temporal characterization of the tipping process. 

The window size needs to be larger than the forcing frequency and smaller than the duration of the tipping process. 
In the C-H  transition, the tipping process lasts nearly 1000~yr, therefore if we want to track the network behaviour during the transition, we need to generate a new network at least ten times. Thus, 
we test the following window sizes: 20, 50 and 100~yr. 
In our case, we are dealing with temporal series with yearly frequency. An analogous argument holds for time series with monthly, daily or hourly frequency, although in this case the daily and seasonal periodicity have to be properly taken into account when calculating the temporal correlations. 

In \fref{Window_size}, we see that the network indicators show a similar behaviour for the three investigated window sizes. For all the indicators, the window size of 100~yr gives a stronger signal than the others. The window size of 20~yr gives a noisy evolution of the network behaviour, sometimes without a clear signal when crossing the tipping point. 
This behaviour can be explained by information theory, since short time series provide less information than longer ones and have higher uncertainty~\cite{Wu2024}.  
Therefore, it is better to work with longer time series. 
However, in the search for early warning signals, we need to have enough points to detect a change of network behaviour. 
The compromise is to 
choose a window size of 50 yr which is sufficient to generate relevant statistics and is 5\% of the tipping transition. This is the value that we will use from now on. 

\subsubsection{Discretization of the forcing rate}

\begin{figure}[t!]  
\includegraphics[width=0.49\textwidth, keepaspectratio]{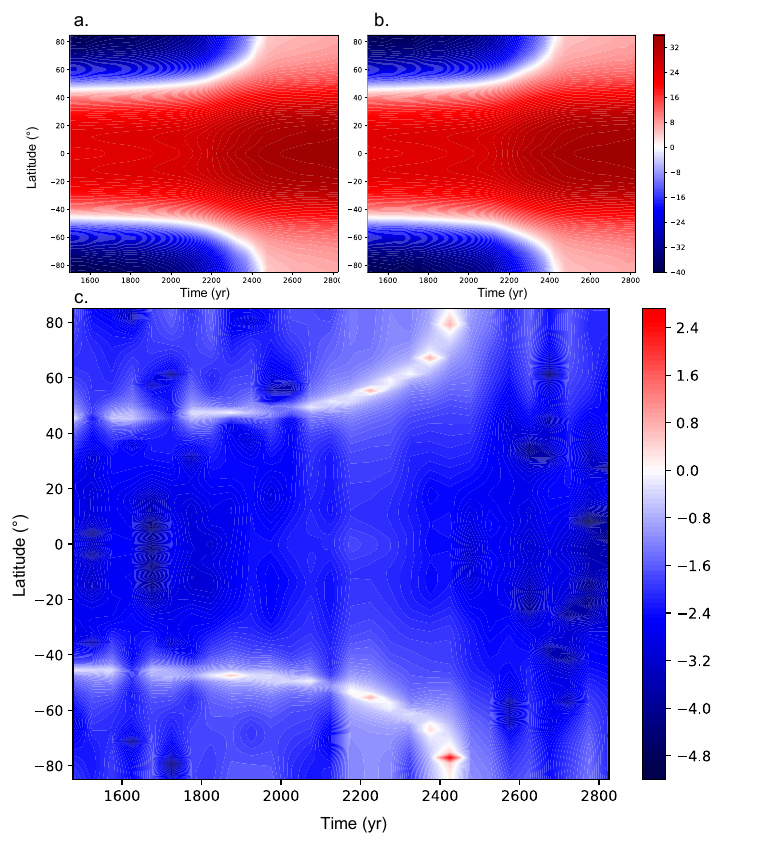}
\caption{Temporal evolution of the zonal average of SAT for an increase of CO$_2$ concentration of (a) 2~ppm each 20~yr; (b) 0.1~ppm per year. (c) Difference between (a) and (b). 
}
\label{Diff_strenght}
\end{figure} 

\begin{figure*}[t!]   
\includegraphics[width=1.0\textwidth, keepaspectratio]{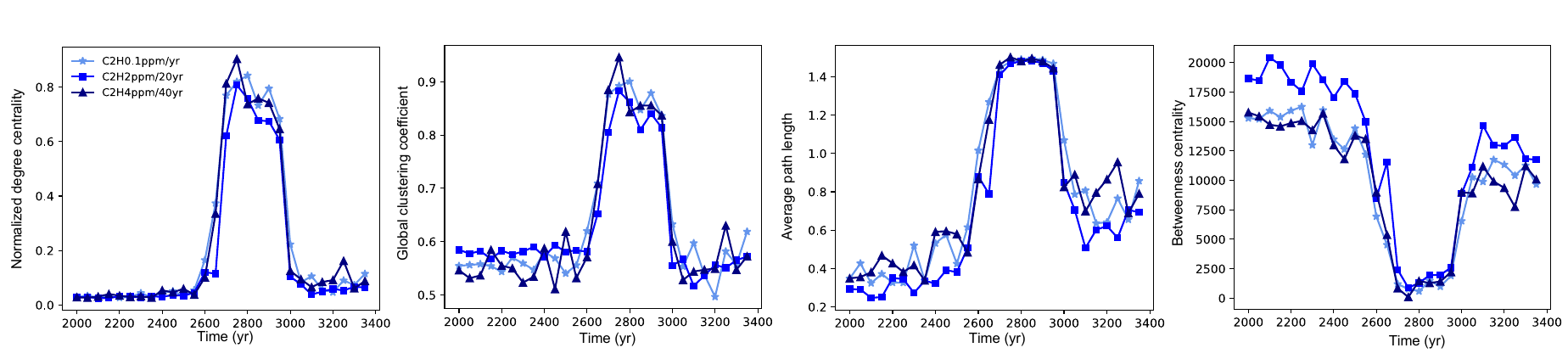}
\caption{Temporal evolution of four network indicators in simulations with different discretization of the forcing rate: 0.1~ppm per year, 2~ppm each 20~yr, and 4~ppm each 40~yr. }
\label{Parameter_strenght_forcing}
\end{figure*} 
 
 There are two types of network stress: internal stress and environmental stress~\cite{Network_resilience}. The internal stress, which depends on the change of the number of vertices, does not occur in our climate networks by construction.  
 On the other hand, the environmental stress acts on the spatial patterns of the network.  
As said before, the transition from one state to another is induced by an increase or decrease of the  external forcing, in our case the atmospheric CO$_2$ content. Thus, we use this forcing as environmental stress, 
by considering the same average rate of 0.1 ppm/yr but with different temporal discretizations, 
0.1~ppm every year, 2~ppm every 20~yr and 4~ppm every 40~yr, starting and ending at the same CO$_2$ concentration values. 
We focus the analysis around the tipping.

To understand how the environmental stress acts on the network we first need to verify if it affects the state variable that is extracted from the climate model. 
In Fig.~\ref{Diff_strenght}, we compare the temporal evolution of SAT zonal averages for coarse and fine discretizations of the forcing. There are indeed localised significant differences between the cases, showing that the climate model is sensitive to the change of the forcing step. Now, we need to understand how this sensitivity affects spatial correlations.
Fig.~\ref{Parameter_strenght_forcing} shows the evolution of the four considered network indicators, defined in Sec.~\ref{subsec:indicators}. 
It turns out that
the overall behaviour of these indicators is similar and quite robust with respect to the change of the forcing step.
The small discrepancies in the indicators are just unveiling the network sensitivity to the fact that in the case of coarse discretization the forcing is stronger, so that the system is farther away from the equilibrium than with fine discretization. 

From now on, we will use yearly forcing steps.

\subsubsection{Forcing rates}

\begin{figure*}   
\includegraphics[width=\textwidth, keepaspectratio]{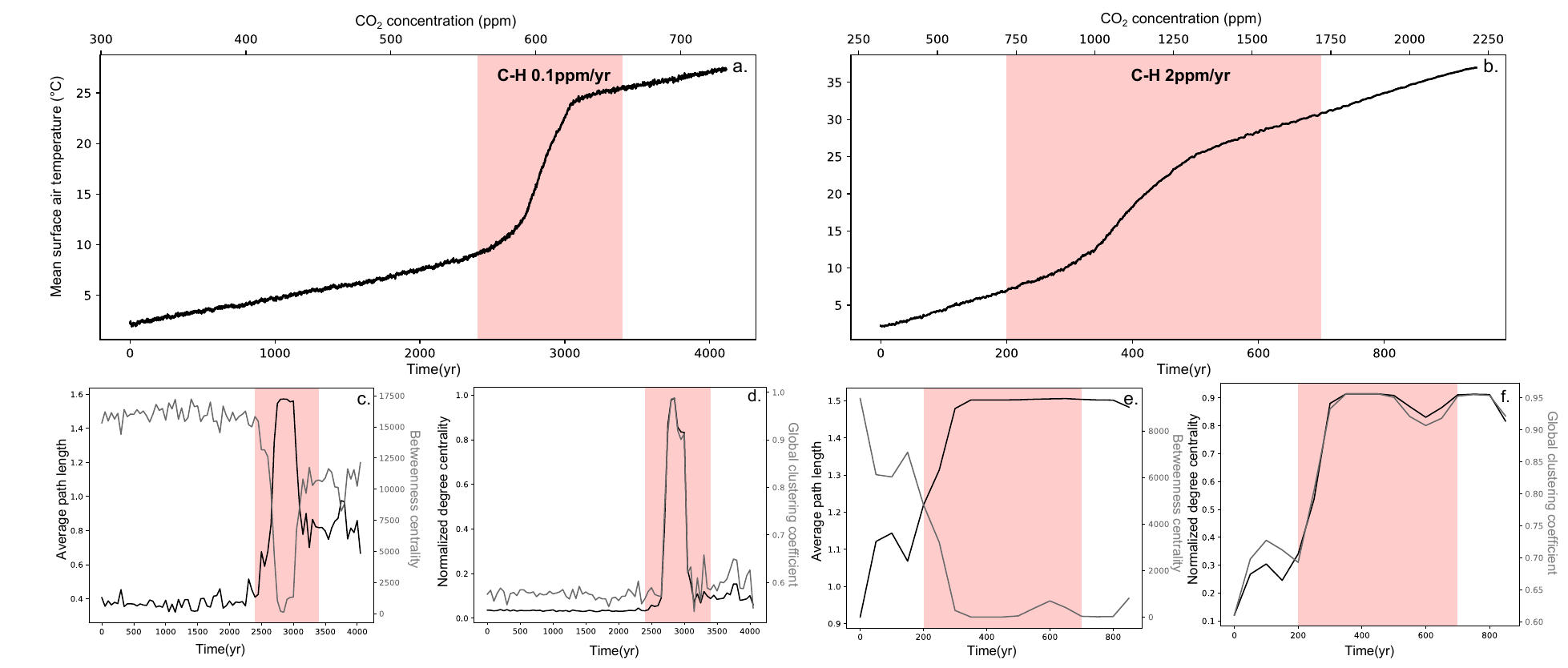}
\caption{Comparison of the two forcing rates. The shaded region corresponds to the tipping interval from the cold state to the hot state for a forcing rate of 0.1~ppm/yr (panels (a), (c), (d)) and of 2~ppm/yr (panels (b), (e), (f)). }
\label{Difference_force}
\end{figure*}

For the C-H transition we compare the two following forcing rates: 0.1~ppm/yr and 2~ppm/yr. The first is sufficiently low to guarantee that the system follows the stable branch of the cold state until a bifurcation-induced tipping occurs, 
while the second corresponds to the present-day forcing rate and can give rise to rate-induced tipping~\citep{ashwin2017,Hoyer2021,feudel2023,ritchie2023} in the coupled-aquaplanet configuration~\citep{Ragon_2023}. 
Indeed, we verified that 
on the cold attractor (first 200~yr) the mean global SAT is $T_{0} = (2.3 \pm 0.1)~^\circ$C without forcing, $T_\textrm{0.1~ppm/yr} = (2.4\pm 0.2)~^\circ$C for the low forcing rate and $T_\textrm{2~ppm/yr} = (4.4 \pm 1.5)~^\circ$C for the high forcing rate. The first two values are the same within the confidence interval, showing that the system has time to relax back to the attractor under the low forcing rate~\citep{tel2020}.

As shown on \fref{Difference_force}a-b, the SAT evolution is radically different. Starting from the same initial condition, the attractor lasts 200~yr for the 2~ppm/yr forcing rate, whereas it lasts 2400~yr for the 0.1~ppm/yr case. 
This has of course a direct consequence on the network indicators, which are not constant in the cold-attractor region, as shown in panels (e)-(f). 
In a general manner, we see that the behaviour of the networks is similar and the indicators increase/decrease during the tipping process. 
While the latter starts for a CO$_2$ content of 560~ppm when the rate is 0.1~ppm/yr, it is delayed to around 720~ppm for the 2~ppm/yr rate, thus overshooting the bifurcation-induced tipping that can be seen in Fig.~\ref{Diagram_bifurc}. 
Hence, even if the tipping mechanism can be different in the two cases (bifurcation-induced \textit{vs.} rate-induced tipping for low or high forcing rates, respectively) the forcing rate does not have a major impact on the network, which is in both cases reacting to the tipping process (see bottom row of Fig.~\ref{Difference_force}).

\subsection{Global and zonal networks}

After the sensitivity tests, we describe now the behavior of the networks constructed over the entire numerical grid and over limited zonal areas during the C-H and C-S transitions.

The aquaplanet is divided in five latitudinal strips (covering the entire longitudinal range), namely the North and South polar regions (with latitude ranging in $\phi \in (66^\circ, 90^\circ)$ and $(-66^\circ, -90^\circ)$, respectively), the mid-latitude regions 
($\phi \in (23^\circ,66^\circ)$ and 
($-23^\circ, -66^\circ)$) and the Equatorial region ($\phi \in (-23^\circ, 23^\circ)$), 
so that the zonal behaviour of the networks can be investigated.
The aim is to investigate if networks can identify dominant dynamical regions during the stable phase on an attractor and during the tipping process.
The network generation is thus performed by giving as input variables to {\tt pyUnicorn} their values in a specific region (note that there is no significant difference if we generate the network globally and then divide it in regions).

As the aquaplanet is symmetric between the northern and the southern hemispheres, we expect the network to behave in a symmetrical way with respect to the Equator. 
This can be verified by comparing the two mid-latitudinal and the two polar regions. There is indeed a symmetry between the indicators for the two hemispheres, as shown in App.~\ref{app:symmetry}. Besides unveiling the robustness of the networks, comparing the evolution of network indicators for the corresponding zones of both hemispheres bears information on how the internal variability of the simulations translates onto the networks.

\begin{figure*}   
\includegraphics[width=1.0\textwidth, keepaspectratio]{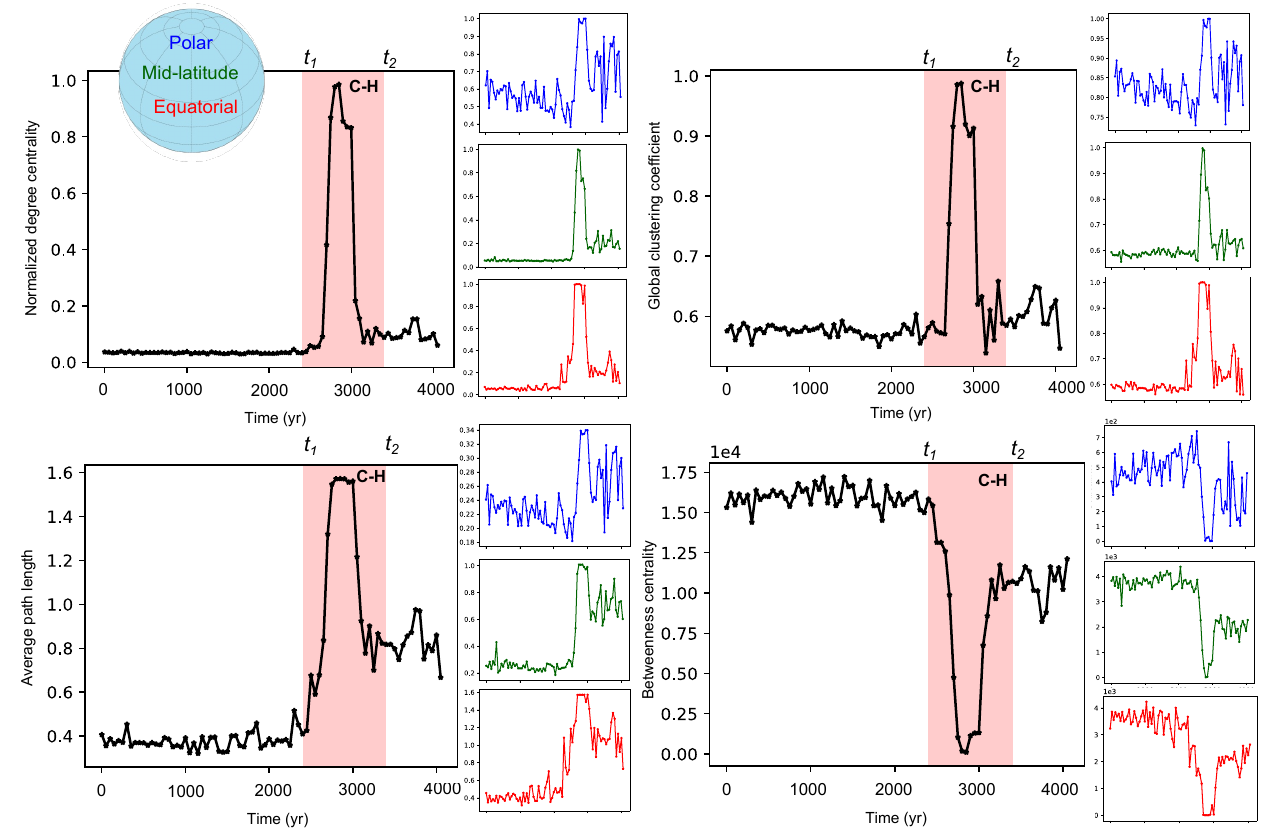}
\caption{Temporal evolution of the four network indicators for SAT in the C-H transition at a rate of 0.1ppm/yr, at global and zonal scales. 
}
\label{C2H_image_ZG}
\end{figure*} 

\begin{figure*}   
\includegraphics[width=1.0\textwidth, keepaspectratio]{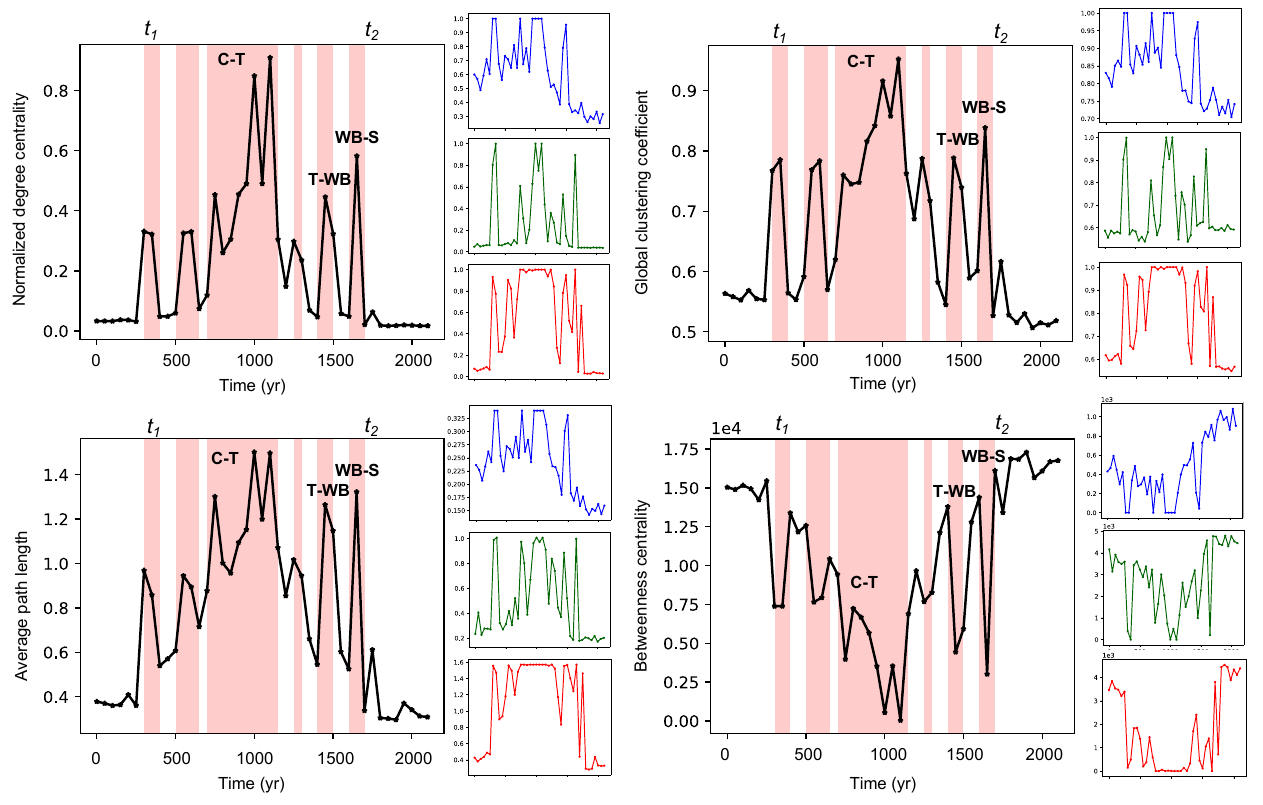}
\caption{Temporal evolution of the four network indicators for SAT in the C-S transition at a rate of 0.1~ppm/yr, at global and zonal scales.}
\label{C2Snow_image_ZG}
\end{figure*} 

\subsubsection{Cold to hot (C-H) transition}

As shown on \fref{C2H_image_ZG} (main panels), the evolution of the network indicators of the C-H transition displays a strong and unique signal during the tipping. For the normalized degree centrality, the global clustering coefficient and the average length distance, a strong increase close to and during the crossing of the tipping process is observed. This is consistent with the previously mentioned {\it slaving principle} near a bifurcation point, with spatial correlations occurring over the whole system~\citep{Haken2004,dakos2010}. 
The behaviour of the average length distance is expected since, as the system transitions, the vertices tend to correlate over longer distances. 
Conversely, the betweenness centrality is strongly decreasing and is approaching zero when the system tips. 
This is also expected, as the vertices correlate over longer distances so that the normalized degree centrality approaches 1.  As all pairs of points tend to be connected, the possibility of alternative paths vanish.

The normalized degree centrality and the global clustering coefficient show higher values in the polar regions on the cold attractor, and they increase in all the zonal regions during the tipping process. 
The equatorial region shows an earlier signal of transition. 
This is likely due to the main dynamical feature in hot-climate attractors~\citep{Brunetti_2019}, where high surface temperatures and the consequent large water vapour-holding capacity of air can initiate deep convection cells, with latent heat released by the condensation of large amounts of moist, ascending air~\citep{Karnauskas2020}. 
The earlier signal of transition in the equatorial region is indeed also observed when specific humidity is used as state variable to construct the networks, as shown in App.~\ref{app:zonal}. 
The same strong signature in the equatorial region is observed in the average length distance.
Finally, the intensity of the internal variability on the cold attractor of the betweenness centrality is smaller at mid-latitudes.  

The evolution of the network indicators in the zonal regions unveils the  importance of the local dynamics in climate. Some regions respond earlier to tipping than others, and show different degrees of internal variability, depending on the attractor.  

\subsubsection{Cold to snowball (C-S) transition}
\label{subsec:C-Sregions}

This transition is shown in Fig.~\ref{Figure_explique}, right panels. Starting from the cold attractor and under a reduction of the CO$_2$ content at a rate of 0.1~ppm/yr, the sea ice undergoes two main episodes of localised disruption and reconstruction at the ice edge at mid-latitudes, as can be seen from the simulation results (not shown). The mean SAT responds to these episodes through the ice-albedo feedback on a timescale of some decades~\citep{Brunetti_2019}. Afterwards, the climate system initiates the longest tipping from the cold to a transient state (denoted as C-T) at around $-20~^\circ$C. After a shift to the waterbelt (T-WB), the climate system finally stabilises to the snowball state (WB-S), where the ocean surface is completely covered by ice. 

As shown in \fref{C2Snow_image_ZG}, the networks indicators react to the fluctuations before the C-T tipping, showing their high sensitivity to local spatial features. In a stronger manner, they react during the C-T, T-WB and WB-S tipping. 
The fact that the networks are sensitive to
a transition that is not as clear as C-H unveils their reliability. This is especially relevant in the case of the Earth climate, where we do not necessarily expect direct transitions as C-H because of the presence of several interacting tipping elements. Moreover, our results show that networks can characterize both upward and downward transitions.

At the beginning, the betweenness centrality shows lower values in the polar region with respect to the other regions, while the other three  network indicators show larger values, showing that active dynamical features are dominant at high latitudes where the ice area is increasing. 
Also the equatorial region is showing a strong signal during the main transition from the cold to the transient state (C-T) as it was the case for the C-H simulation.
However, contrary to the C-H transition, the behavior of the equatorial region in the C-T transition is  driven by the formation of the ice while it extends towards low latitudes. Therefore, we will consider the ice extent as the area where EWS are estimated in Sec.~\ref{subsec:C-Sews}. 

\subsection{Early warning signals (EWS)}

We seek for statistical network indicators that are able to anticipate the transition by giving a clear and non-ambiguous signal. For this, we are going to use our regional study as it has unveiled that some regions are dynamically active and dominant during the transitions, thus  more sensitive than the others. For SAT, the dominant region during the transition appears to be the equatorial region and the polar region for the cloud cover fraction, as shown in  App.~\ref{app:zonal}. Therefore, the idea is to use the temporal evolution of the network indicators in the dominant region for each of the climate state variables, and compare it to 
traditional EWS methods~\citep{Dakos2008,scheffer2009ews,dakos2012,xie2019} such as the variance, the  autocorrelation at lag-1 and the kurtosis. 

These three classical indicators are estimated in the same region as the network ones. However, one has to keep in mind that the information of the regional dominance was extracted from the behaviour of the complex network and not from a classical time series analysis. 
Here, we perform the EWS analysis for the two transitions, C-H and C-S.  
We are going to detail the results only for SAT, while the analysis made for the cloud cover fraction and the relative humidity provide similar results (not shown). 


\subsubsection{Cold to hot (C-H) transition}
\label{subsec:C-Hews}

\begin{figure*}   
\includegraphics[width=1.0\textwidth, keepaspectratio]{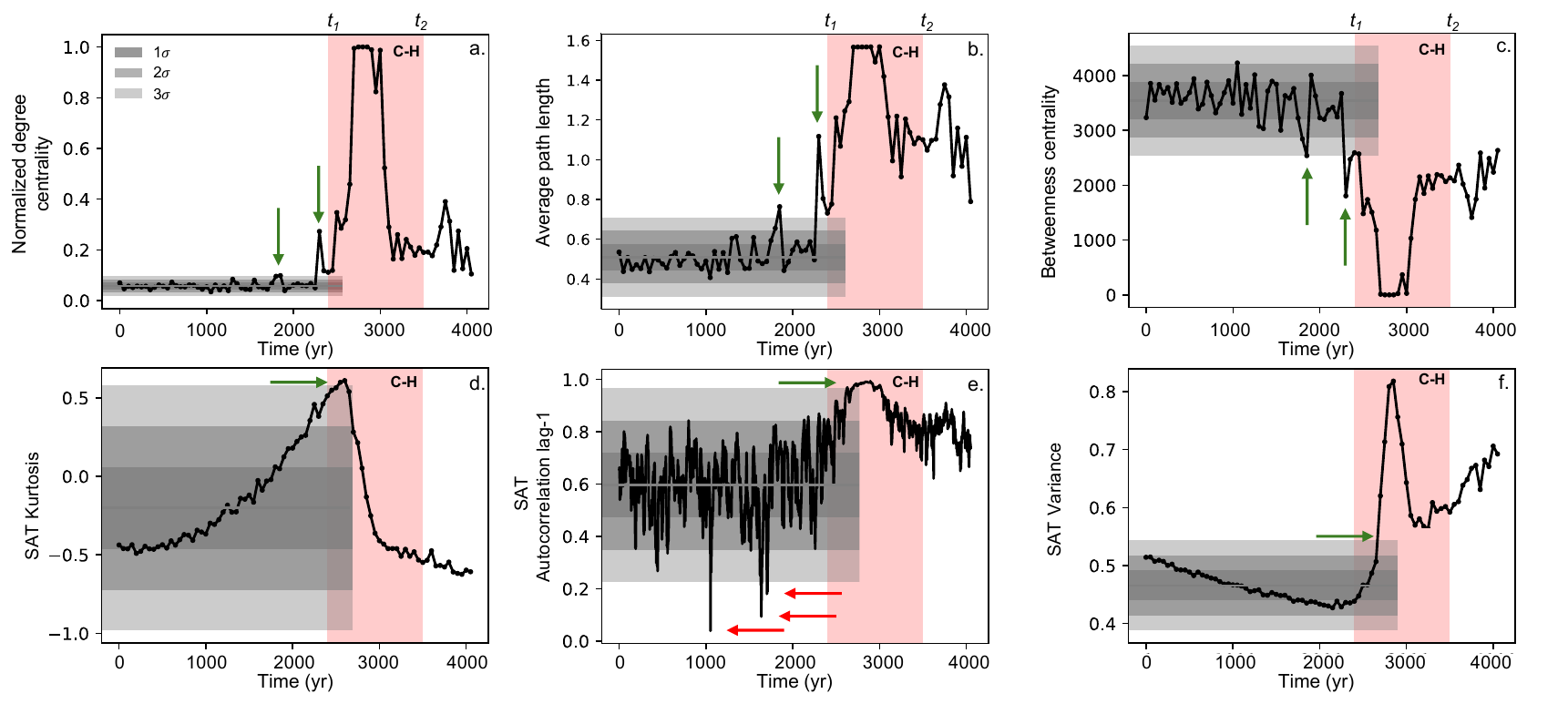}
\caption{Comparison of six EWS indicators in the equatorial region, during the C-H transition at a rate of 0.1~ppm/yr. a., b. and c. are network based and d., e. and f. are time series based.}
\label{EWS_eqsigmaC2H}
\end{figure*} 

\fref{EWS_eqsigmaC2H} shows differences in the behaviour of the candidate EWS indicators. 
On the plots the 1$\sigma$, 2$\sigma$ and 3$\sigma$ intervals are labelled assuming a normal distribution of each indicator on the cold attractor. A point is identified as an EWS when it overcomes the 3$\sigma$ interval.
This strict limitation gives a confidence of 
99.73\% in the signal.  

For the three network-based methods (panels (a), (b), (c)), we see that the EWS occurs at 1850~yr for the normalized degree centrality, the average length distance and the betweenness centrality, whereas the kurtosis gives a signal of EWS at 2550 years (panel (d)). The autocorrelation at lag-1 (panel (e)), which should increase near a tipping point~\citep{dakos2012}, gives a signal at 2650~yr, with three contradictory low-level signals labelled by the red arrows. Also, we notice that the 3$\sigma$ interval takes 75\% of the panel (e), due to the high internal variability on the cold attractor, which produces negative spikes exceeding $3\sigma$ in the autocorrelation at lag-1, challenging the normality assumption in that case. 
Finally, the variance decreases on the attractor and becomes an EWS at 2700~yr. 

The three network indicators give comparable results, even if the ones using global topological information, such as the average length distance and the betweenness centrality, turn out to be more noisy   than the normalized degree centrality, which is a local indicator.
In summary, networks-based indicators, which take into account both spatial and temporal features of the dynamical process, perform better as EWS than the traditional ones. We also note that kurtosis and variance drift substantially while the system is still well on the attractor, which artificially widens their standard deviations. Conversely, the network indicators are remarkably stable over the attractor, which facilitates the definition of their typical behaviour and eases the detection of the sudden deviations at the approach of the tipping.

\subsubsection{Cold to snowball (C-S) transition}
\label{subsec:C-Sews}

We perform a similar analysis for the C-S transition. In this case, we focus on the ice formation, since we saw in Sec.~\ref{subsec:C-Sregions} that this is the main dynamical process that influences the overall dynamics. Therefore, the region where we construct the networks  is the ice extent area, which evolves in time. The traditional indicators are also estimated in this region.

\begin{figure*}   
\includegraphics[width=1.0\textwidth, keepaspectratio]{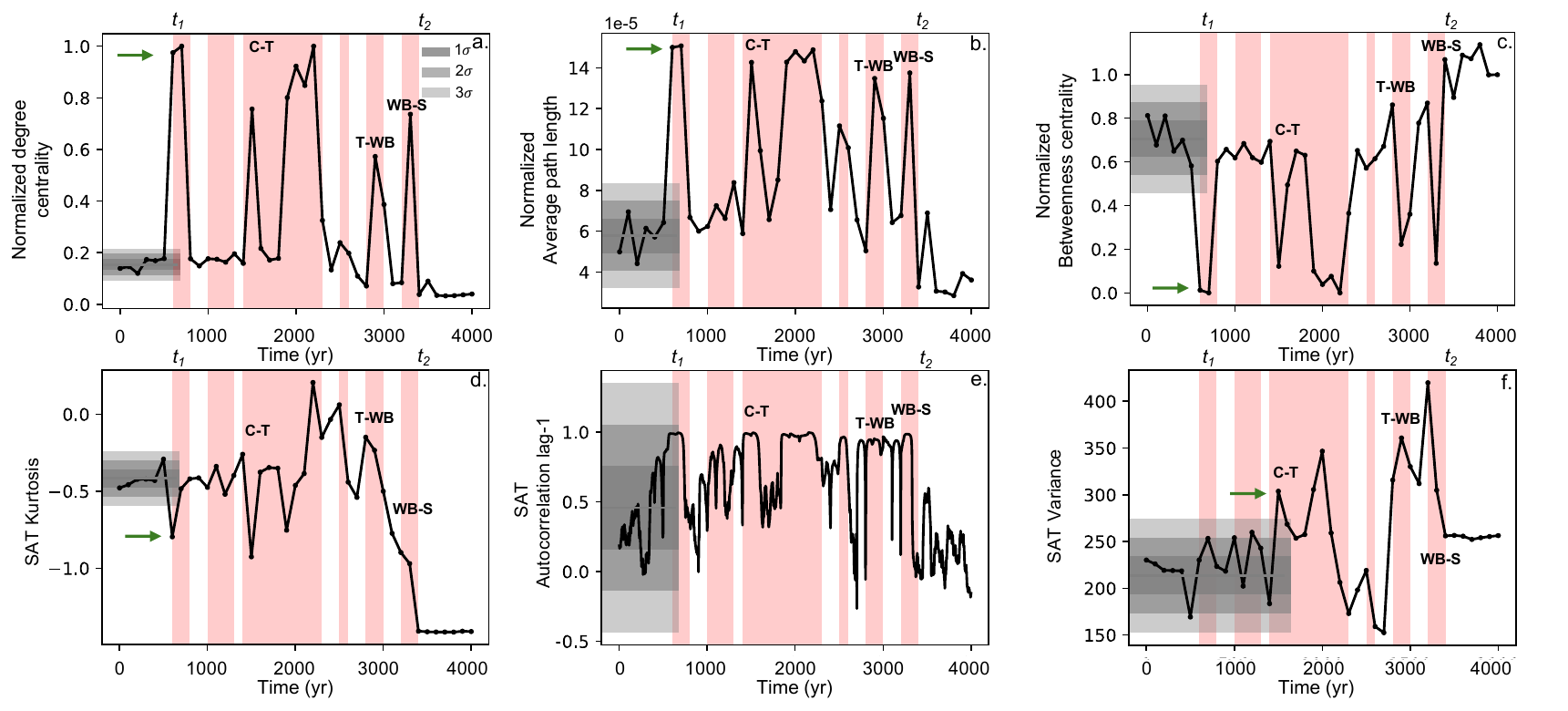}
\caption{Comparison of six EWS in the ice-extent region during the C-S transition at a rate of 0.1~ppm/yr, where a. and  b. are network based and d., e. and f. are time series based.}
\label{EWS_C2S}
\end{figure*}

As shown in \fref{EWS_C2S}, the network-based indicators and the kurtosis to a certain extent detect at 300~yr the ice-disruption episodes before the main tipping (C-T). As already remarked in the case of the C-H transition (Sec.~\ref{subsec:C-Hews}), the strong fluctuations in the autocorrelation at lag-1 do not enable us to see the EWS. 
 These fluctuations are an artifact due to an extreme temporal stability of the zonally-averaged SAT, within $\pm 0.01^\circ$C Therefore, the autocorrelation at lag-1 is mostly measuring random noise and is irrelevant. In contrast, during the tipping the fluctuations of the SAT increase to 0.1$^\circ$C and become physically meaningful, so that the autocorrelation becomes relevant and its value stabilises tending to 1.

Finally, the standard deviation shows a signal at 750~yr during the C-T tipping, with strong oscillations within the 3$\sigma$ interval during the abrupt episodes of ice disruption.
However, the fast succession of stable and tipping intervals in the C-S transition prevents an EWS to display clearly for each tipping process.

\subsection{Non-linear climate dynamics}

An additional advantage of climate networks is that they allow to find where the non-linear dynamics is taking place in the climate system. This can be understood by identifying the non-linear edges in the climate network. To do so, we generate a network with PCC to obtain the linear edges, and with MI to obtain the non-linear edges. Then, to perform the difference between these two networks in order to keep the non-linear part only, we use the method described in \citet{Donges_2009}. We compare the outcomes at two different time slices, on the cold attractor and during the tipping process of the C-H transition. 

A necessary condition that arises to do this comparison is to have the same edge density when generating the networks with PCC and MI~\citep{Donges_2009}.  
Thus, as in all the above analysis, we set the PCC threshold to $0.5$ during the tipping. At 2650~yr (see the evolution plot in Fig.~\ref{Figure_explique}a), this corresponds to an edge density of 0.05. The same edge density is obtained in the MI network with a  threshold of $0.29$. 
Since we showed in Sec.~\ref{subsec:threshold} that the PCC threshold during the tipping is not saturated at 0.5, we keep an edge density of 0.05 for all the generated networks
at both time slices, in the cold attractor (at $t=$50~yr) and during the tipping (at $t = 2650$~yr), as summarised in Table~\ref{NLEvaleur}.

\begin{table}[]
    \centering
    \begin{tabular}{c|cc|cc}
      &  \multicolumn{2}{c|}{ Cold attractor} & \multicolumn{2}{c}{ Tipping}  \\
      & PCC & MI & PCC & MI  \\
     \hline
Threshold & 0.10 & 0.03 & 0.5 & 0.29  \\
Edge density & 0.05 & 0.05 & 0.05 & 0.05  \\
    \end{tabular}
    \caption{Threshold and edge density values for the cold attractor (50~yr) and the tipping (2650~yr). 
    }
    \label{NLEvaleur}
\end{table}

\begin{figure*}   
\includegraphics[width=1.0\textwidth, keepaspectratio]{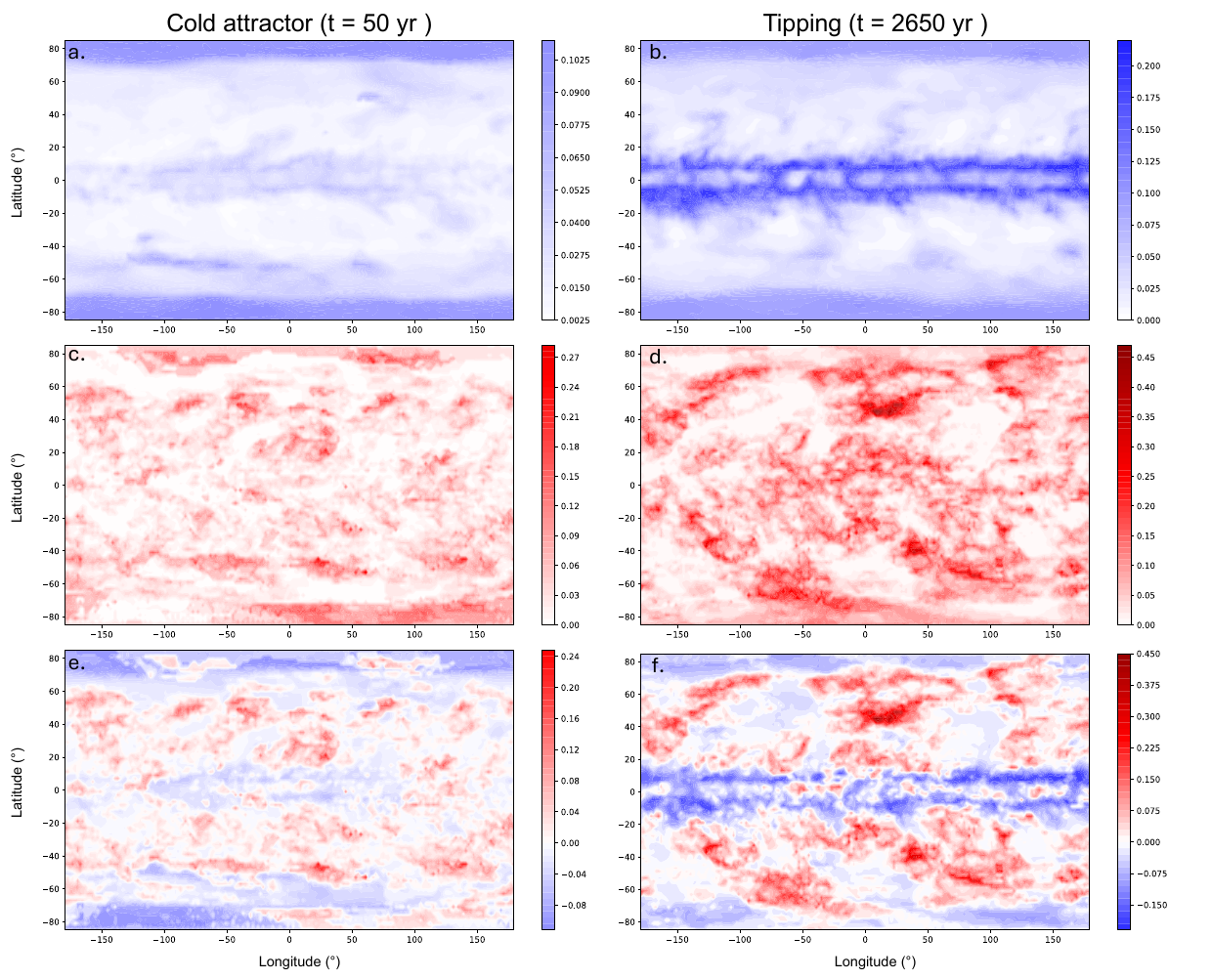}
\caption{First column is in the cold attractor region (50~yr) with a. PCC, c. MI and e. the difference between MI and PCC. Second column corresponds to the tipping (2650~yr) with b. PCC, d. MI and f. the difference between MI and PCC.}
\label{NLE_image}
\end{figure*}

In Fig.~\ref{NLE_image}, we show the edges as density maps for PCC and MI networks, respectively, as well as their difference, at the two time slices. 
We see that on the cold attractor 
(panels in the left column), the strongest linear correlations occur in the polar region, while non-linear ones are quite homogeneously occurring over all the planet. In contrast, during the tipping process (panels in the right column), the strongest linear correlations are due to the propagation of Kelvin waves in the equatorial region, while the patterns in the non-linear correlations correspond to the propagation of atmospheric Rossby waves, which transmit circulation anomalies in the surface temperature from the tropical region into the midlatitudes in each hemisphere~\citep{hoskins1981,trenberth1998,Karnauskas2020}. These linear (in blue) and non-linear (in red) features are clearly evident in the difference map (panel (f)). The behaviour of climate networks helps evidencing physical teleconnections at large scales, which become dominant during the tipping process.  

\section{Discussion and conclusions}

In summary, using coupled-aquaplanet simulations showing a global-scale tipping from a climatic attractor to another, we have verified that network indicators have much better properties as early warning signals (EWS) of tipping processes than traditional ones. This is due to the fact that they encode both spatial and temporal information. 

Moreover, they show high sensitivity to local dynamical features, like sea ice formation or disruption, which enables to use network indicators also for tipping occurring at the regional level. This is relevant, since some tipping elements in the climate system~\citep{McKay2022}, like the Amazon rainforest dieback or the collapse of Greenland and West Antarctic ice sheets, can first have an impact on the subcontinental scale before affecting the fonctioning of the whole climate at the global scale through tipping cascade~\citep{niko2024}.

We have tested that the ability of network indicators as EWS persists in systems forced at high rates with an overshoot of the bifurcation-induced tipping point. Beside rate-induced tipping, other tipping mechanisms can develop in the climate system, such as noise- or shock-induced tipping~\citep{ashwin2012,Ragon_2023}. These occur when the internal variability of the attractor dynamics  reaches a critical threshold close to the basin boundary between alternative attractors, and can be triggered, for example, by perturbations in the volcanic activity~\citep{baum2022} or in the biological pump~\citep{MUKHOPADHYAY2008,elke2020}. Finding characteristic signatures in the behavior of network indicators and topology during tipping processes of different nature requires further investigation.  

Finally, an additional advantage of complex networks is that alternative ways of network construction permit to isolate the non-linear dynamical components. In the case of coupled-aquaplanet, we have seen that the main nonlinear feature on the interannual time scale is the propagation of Rossby waves from the Equator to mid-latitudes~\citep{trenberth1998}. In general, this type of analysis can be used to characterise the backbone structure of each attractor and the main teleconnections taking place on each attractor and during the tipping 
process. Using time series with daily or monthly frequencies, climatic phenomena with different characteristic timescales can be investigated using, for example, the so-called ordinal analysis~\citep{PhysRevLett.88.174102,deza2015}.       

Considering the huge current production of satellite data, methods that exploit both their spatial and temporal coverages, like those for network constructions and indicators, become of particular interest. We plan to apply such methods to present-day climate simulations participating to the Tipping Point Modelling Intercomparison  Project (TIPMIP), which is presently defining its protocols and analysis procedures.   


%
%

%

\begin{acknowledgments}
We thank Cristina Masoller for useful discussions. We acknowledge the financial support from the Swiss National Science Foundation (Sinergia
Project No.~CRSII5\_213539).
\end{acknowledgments}

\section*{Author declarations}

\subsection*{Conflict of Interest Statement} 

The authors have no conflicts to disclose.

\subsection*{Author Contributions}

{\bf Laure Moinat}: conceptualization (supporting); formal analysis (equal); visualization (lead); methodology (equal); writing/original draft (equal); writing/review and editing (equal). 
{\bf J\'er\^ome Kasparian}: methodology (equal); supervision (equal); writing/review and editing (equal).
{\bf Maura Brunetti}: conceptualization (lead); methodology (equal); formal analysis (equal); supervision (equal); writing/original draft (equal); writing/review and editing (equal); funding acquisition (lead).

\section*{Data Availability Statement}

The data that support the findings of
this study and the python scripts are available from the corresponding author upon reasonable request, and will be openly available on a Yareta repository at the University of Geneva~\citep{Yareta_prov} upon acceptance. Data were generated
by the MIT general circulation model (version c67f) and by the python package {\tt pyUnicorn}, which are openly
available at https://github.com/MITgcm and https://www.pik-potsdam.de/\~donges/pyunicorn, respectively.

\bibliography{example}

\section{Appendix}

\subsection{The pyUnicorn package}

The \texttt{pyUnicorn} package in python~\citep{Donges_2015} has the ability to combine the non-linear time series analysis and the complex network methods applied to climate. The package is open source. It can be applied to other fields, and therefore, we are not going to detail all the classes but only the ones that are relevant regarding our analysis. Its main functionality is to import several time series, to generate the complex network with the possibility to use different statistical method for the threshold, and to give in output several files with the different analyses that have been performed. 

Its inner structure is mainly dependent on the \texttt{Numpy}, \texttt{Scipy} and \texttt{igraph} python libraries. The basic network generation is performed using the \texttt{igraph} library. 
More costly computation is done using C, C++ and FORTRAN compilers in \texttt{Cython}~\citep{Behnel2011}. 

This package is divided in five big classes: \texttt{core}, \texttt{functnet}, \texttt{climate}, \texttt{timeseries} and \texttt{utils}. We are going to focus on the \texttt{core} and \texttt{climate} classes as we are only using these ones. The \texttt{core} class is the basic building of the network as it allows one to import data in the \texttt{NetCDF} format using the \texttt{Data} class. The \texttt{climate} class is built on \texttt{GeoNetwork} and \texttt{Data}, allowing one to build networks using different statistical constraints, such as \texttt{climate.TsonisClimateNetwork} and \texttt{climate.MutualInfoClimateNetwork}. The statistical constraints can be either fixed by \texttt{climate.setthreshold} or \texttt{climate.setlinkdensity}.

\subsection{Hemispherical symmetry}
\label{app:symmetry}

\begin{figure}[h!]   
\includegraphics[width=0.49\textwidth, keepaspectratio]{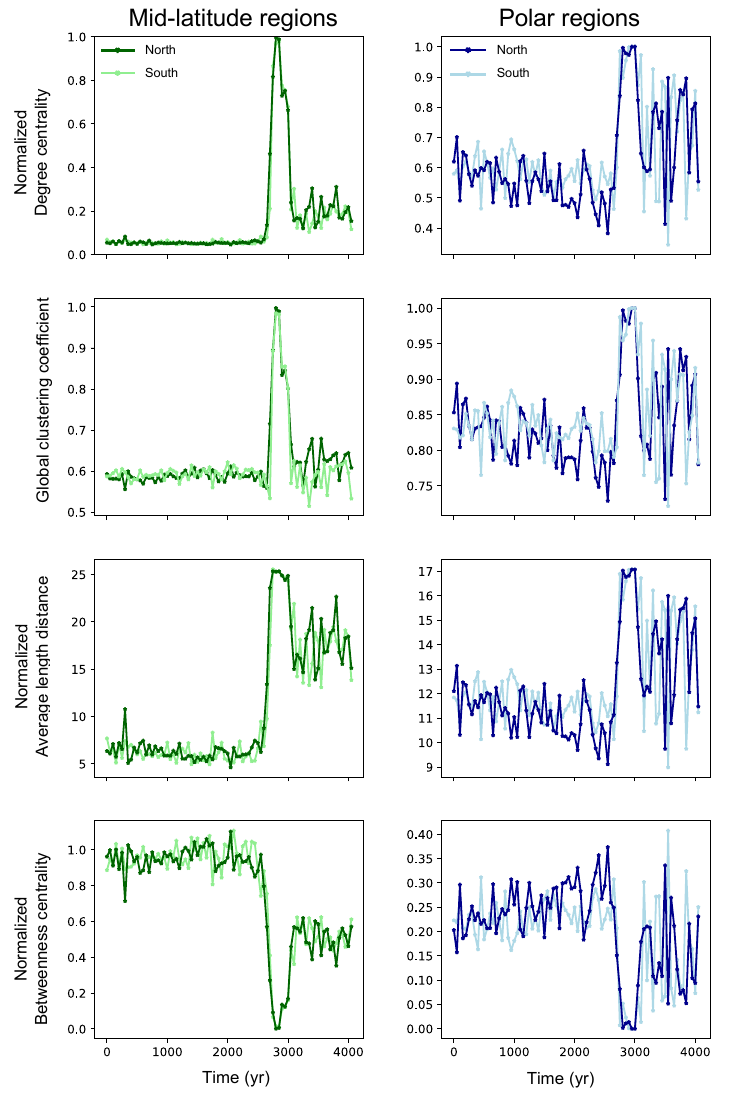}
\caption{Comparison of the evolution of the network indicators for SAT in the two hemispheres in the case of the C-H transition at a rate of 0.1~ppm/yr: 
a. Mid-latitudinal regions and b. Polar regions.}
\label{Symmetry_SAT_C2H}
\end{figure} 

The symmetric behaviour of the network indicators in the northern and southern hemispheres of the aquaplanet can be observed by comparing the two mid-latitudinal and the two polar regions, as shown in Fig.~\ref{Symmetry_SAT_C2H}. The internal variability on the attractor is translated to the small fluctuations of the network indicators, unveiling their robustness.

\subsection{Global and zonal networks for cloud cover and specific humidity}
\label{app:zonal}

\begin{figure*}   
\includegraphics[width=0.9\textwidth, keepaspectratio]{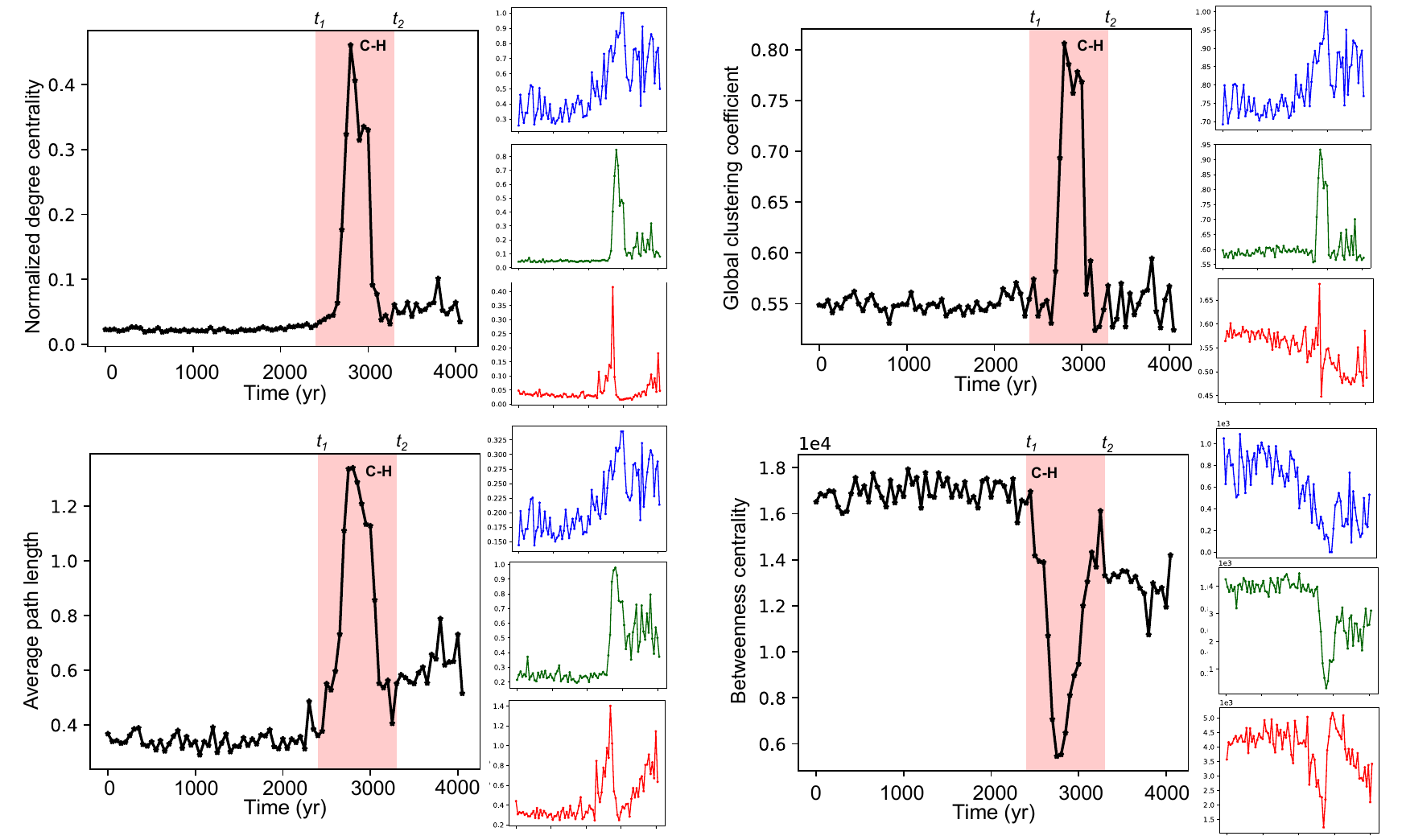}
\caption{Temporal evolution of the four network indicators for specific humidity in the C-H transition at a rate of 0.1~ppm/1yr,  at global and zonal scales. 
}
\label{QS_ZG}
\end{figure*}

\begin{figure*}   
\includegraphics[width=0.9\textwidth, keepaspectratio]{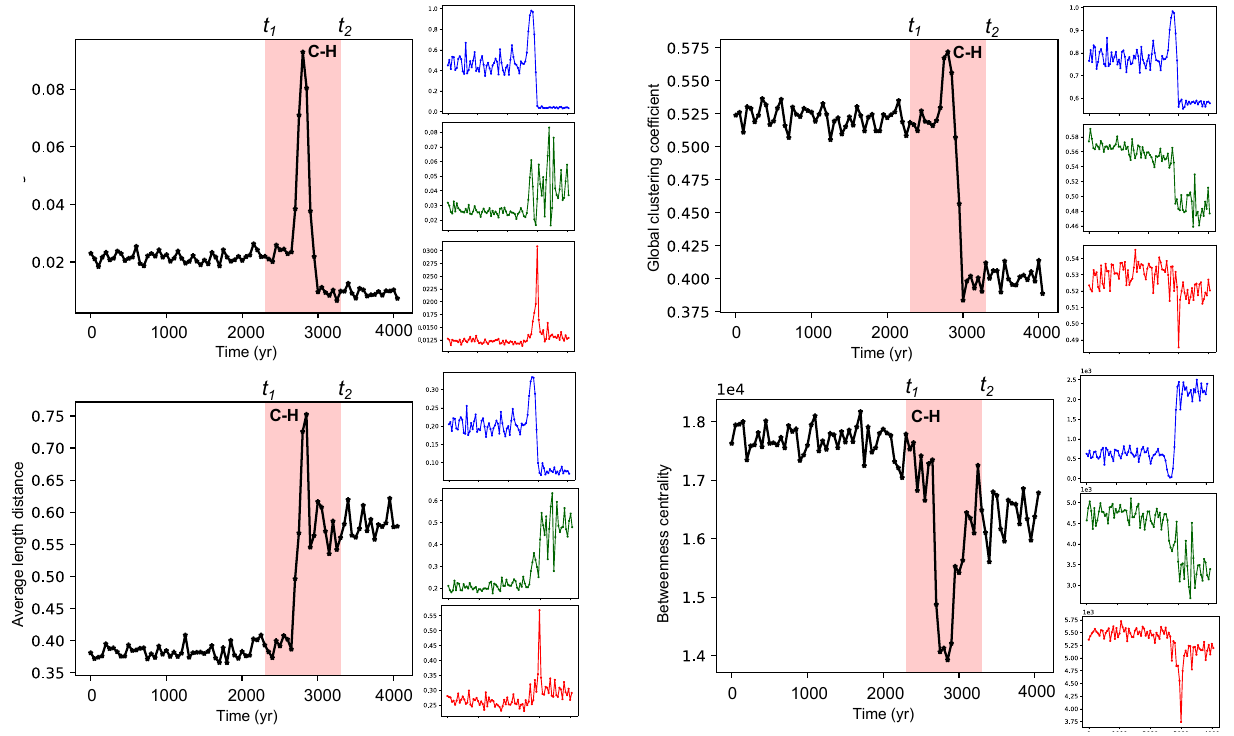}
\caption{Temporal evolution of the four network indicators for cloud cover in the C-H transition at a rate of 0.1~ppm/1yr, at global and zonal scales. 
}
\label{CC_ZG}
\end{figure*}

\begin{figure*}   
\includegraphics[width=0.9\textwidth, keepaspectratio]{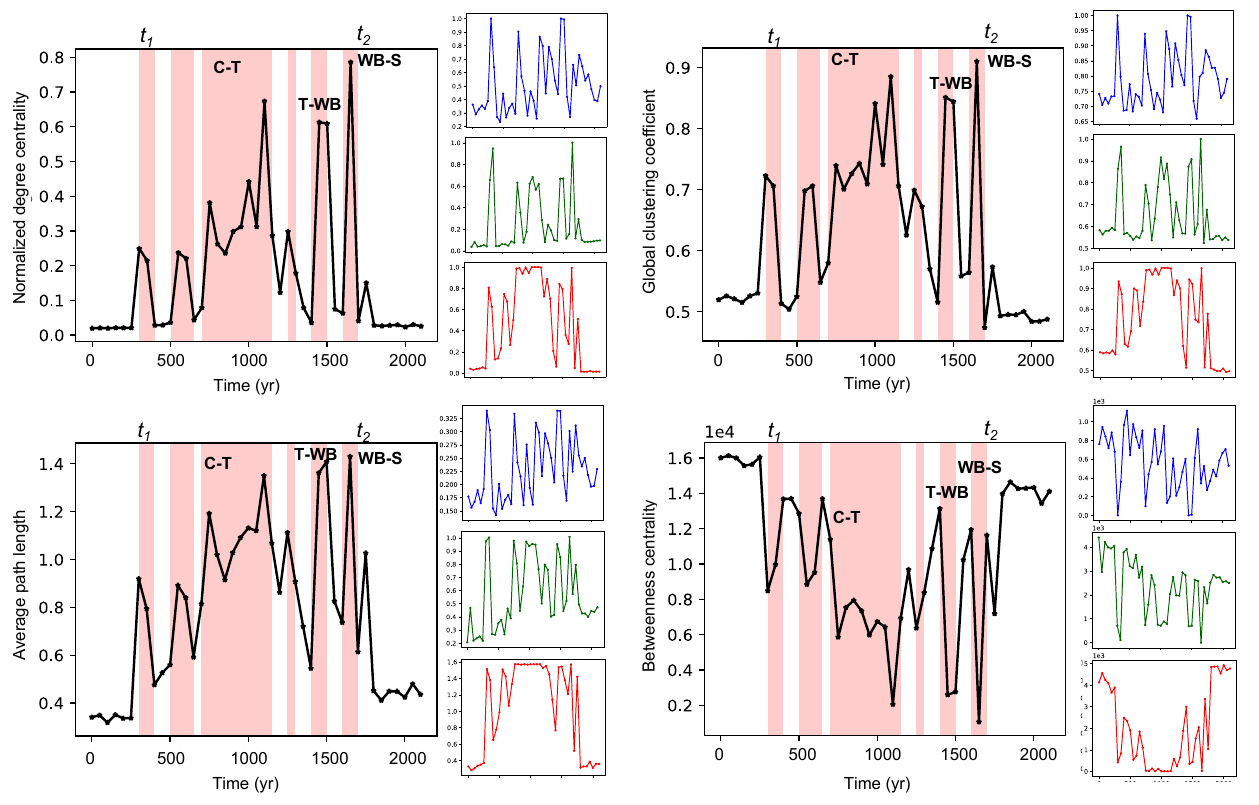}
\caption{Temporal evolution of the four network indicators for specific humidity in the C-S transition at a rate of 0.1~ppm/yr, at global and zonal scales. 
}
\label{C2S_QS_Image}
\end{figure*}

\begin{figure*}   
\includegraphics[width=0.9\textwidth, keepaspectratio]{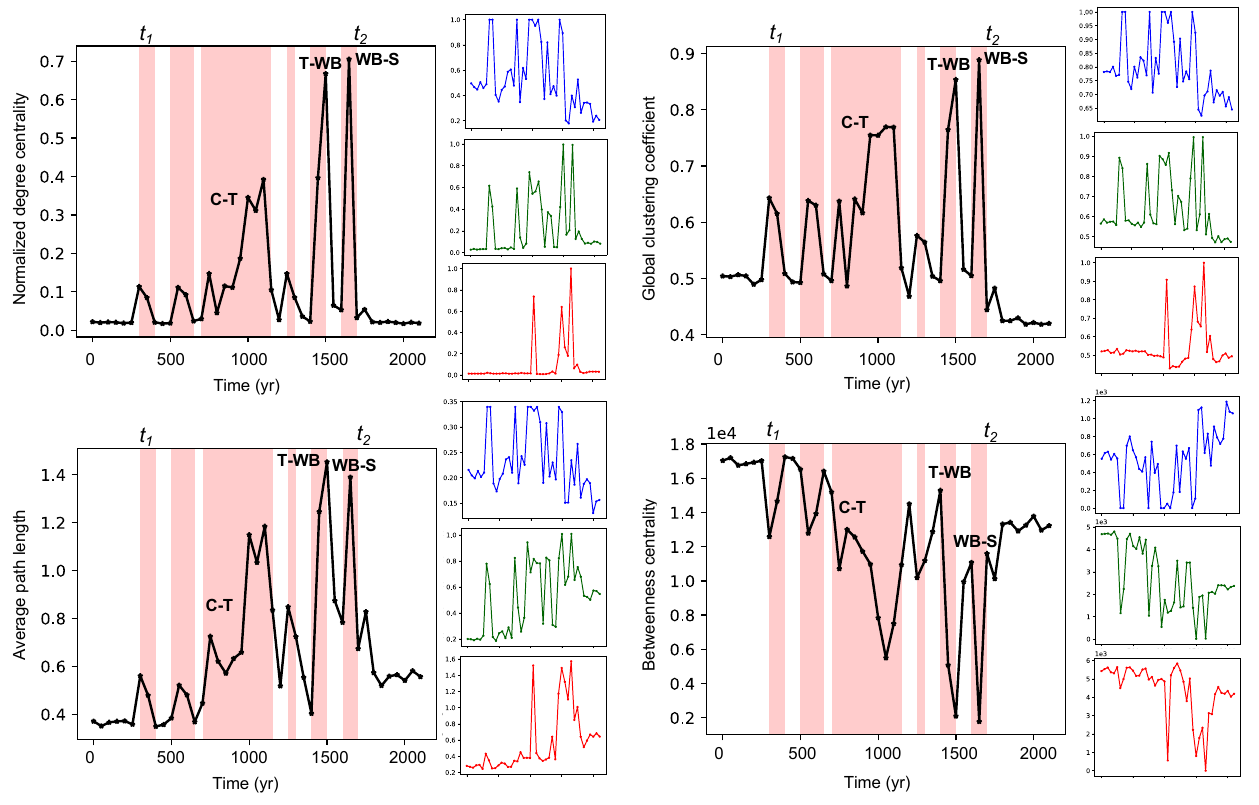}
\caption{Temporal evolution of the four network indicators for cloud cover in the C-S transition at a rate of 0.1~ppm/yr,  at global and zonal scales. 
}
\label{C2S_CC_Image}
\end{figure*}

We report here the behavior of networks constructed from two  state variables complementary to the SAT detailed in the main text: cloud cover and specific humidity. The behaviour of the same four network indicators as in the main text during the 
C-H and C-S transitions is shown in 
Figs.~\ref{QS_ZG}, \ref{CC_ZG}, \ref{C2S_QS_Image}, and \ref{C2S_CC_Image}, which constitute direct counterparts of Figs.~\ref{C2H_image_ZG} and \ref{C2Snow_image_ZG}.
The indicators constructed using cloud cover and specific humidity show a similar behaviour to those based on SAT that we have discussed in the main text, in both the global and zonal regions. Thus, all these variables, which are accessible from satellite data, can be equally used to construct climate networks.

\end{document}